\documentclass[sigconf]{acmart}

\usepackage{booktabs}
\usepackage{multirow}
\usepackage{enumitem}

\def \cU {\mathcal{U}}
\def \cI {\mathcal{I}}
\def \cR {\mathcal{R}}
\def \cA {\mathcal{A}}
\def \cV {\mathcal{V}}
\def \cX {\mathcal{X}}
\def \cB {\mathcal{B}}
\def \bs {\mathbf{s}}
\def \bh {\mathbf{h}}
\AtBeginDocument{%
  \providecommand\BibTeX{{%
    \normalfont B\kern-0.5em{\scshape i\kern-0.25em b}\kern-0.8em\TeX}}}

\copyrightyear{2021}
\acmYear{2021}
\setcopyright{acmcopyright}
\acmConference[WSDM '21]{Proceedings of the Fourteenth ACM International Conference on Web Search and Data Mining}{March 8--12, 2021}{Virtual Event, Israel}
\acmBooktitle{Proceedings of the Fourteenth ACM International Conference on Web Search and Data Mining (WSDM '21), March 8--12, 2021, Virtual Event, Israel}
\acmPrice{15.00}
\acmDOI{10.1145/3437963.3441726}
\acmISBN{978-1-4503-8297-7/21/03}

\renewcommand{\baselinestretch}{0.98}
\begin{document}
\fancyhead{}

\title{Explanation as a Defense of Recommendation}


\author{Aobo Yang$^1$, Nan Wang$^1$, Hongbo Deng$^2$, Hongning Wang$^1$}
\affiliation
{
    \institution{$^{1}$University of Virginia, Charlottesville, USA}
}
\affiliation
{
    \institution{$^{2}$Alibaba Group, Hangzhou, China}
}
\email{{ay6gv, nw6a}@virginia.edu, dhb167148@alibaba-inc.com, hw5x@virginia.edu}

\renewcommand{\shortauthors}{Aobo Yang, Nan Wang, Hongbo Deng, Hongning Wang}

\begin{abstract}
Textual explanations have proved to help improve user satisfaction on machine-made recommendations. However, current mainstream solutions loosely connect the learning of explanation with the learning of recommendation: for example, they are often separately modeled as rating prediction and content generation tasks. In this work, we propose to strengthen their connection by enforcing the idea of sentiment alignment between a recommendation and its corresponding explanation. At training time, the two learning tasks are joined by a latent sentiment vector, which is encoded by the recommendation module and used to make word choices for explanation generation. At both training and inference time, the explanation module is required to generate explanation text that matches sentiment predicted by the recommendation module. Extensive experiments demonstrate our solution outperforms a rich set of baselines in both recommendation and explanation tasks, especially on the improved quality of its generated explanations. More importantly, our user studies confirm our generated explanations help users better recognize the differences between recommended items and understand why an item is recommended.
\end{abstract}



\maketitle

\section{Introduction}

After extensive amount of research effort endeavored to advance the recommendation algorithms \cite{koren2009matrix, he2017neural, sarwar2001item, rendle2010factorization, aggarwal2016recommender}, solutions that explain the machine-made decisions have recently come under the spotlight \cite{herlocker2000explaining,zhang2018explainable}. Numerous studies have shown that explanations help users make more accurate decisions \cite{bilgic2005explaining}, improve their acceptance of recommendations \cite{herlocker2000explaining}, and build up confidence towards the recommender system \cite{sinha2002role}.

Textual explanations have been identified as a preferred medium for explaining the recommendations \cite{zhang2018explainable}, e.g., ``\textit{This restaurant's decoration is unique and its sandwich is the best}''. But due to the lack of explicit training data, most existing solutions appeal to user reviews as a proxy \cite{zhang2014explicit,wang2018explainable,wang2018reinforcement,tao2019the,chen2018neural,truong2019multimodal, sun2020dual}: a good explanation should overlap with user-provided reviews.  
This is backed by extensive prior research in sentiment analysis \cite{DBLP:journals/ftir/PangL07} that there is a strong correlation between opinion ratings and associated review content. 
But the approximation also inadvertently shifts the objective of explanation learning to generating or even memorizing reviews, in a verbatim manner. 
It unfortunately drives the current practice in explainable recommendation to decoupling the learning of recommendation and explanation into two loosely linked sub-problems with their own objectives (e.g., rating prediction vs., content reconstruction) \cite{zhang2014explicit,li2017neural,wang2018explainable}. 
But we have to emphasize that the content generated with fairly fluent language is not sufficient to be qualified explanations, as a good explanation must elaborate why the recommendation is particular to the user. Ideally, based on the provided explanations, a user should reach the same conclusion as the system does about why an item is recommended, i.e., explanation as a defense of the recommendation.

\begin{table*}
  \caption{Case study on two explainable recommendation algorithms' output. Two restaurants are evaluated by the two algorithms, with corresponding recommendation scores and explanation output. We manually labeled attribute words in italic and sentiment words in bold.}
  \vspace{-1mm}
  \label{tab:example}
  \begin{tabular}{|l|lp{7.0cm}|lp{6.8cm}|}
    \hline
    \multirow{2}*{Item} & \multicolumn{2}{l|}{Algorithm 1 (Our proposed model)} & \multicolumn{2}{l|}{Algorithm 2 (NRT \cite{li2017neural})} \\
      &  Score  & Explanation & Score & Explanation \\
    \hline
    {\it A} & 4.2 & the \emph{sushi} is \textbf{good}, the \emph{rolls} are \textbf{fresh} and the \emph{service} is \textbf{excellent}. & 4.1 & their \emph{prices} are \textbf{decent}, but the \emph{portions} are \textbf{pretty small}. \\
    {\it B} & 2.1 & it was a bit \textbf{loud} and the \emph{service} was \textbf{slow}. & 2.2 & \textbf{great} \emph{food}, \textbf{clean}, and \textbf{nice} \emph{atmosphere}. \\
  \hline
\end{tabular}
\vspace{-3mm}
\end{table*}

To tie the loose ends in explainable recommendation, one needs to understand how users perceive and utilize the system-provided explanations. A recent user behavior study based on eye-tracking \cite{chen2019user} finds that opinionated explanations at a detailed attribute-level stimulate users to compare across related recommendations, which in turn significantly increase users' product knowledge, preference certainty, and perceived recommendation transparency and quality. 
Motivated by this finding, we believe the sentiment delivered by the explanation text needs to reveal the details of how items are scored and ranked differently by the system. 
We formulate this as \emph{sentiment alignment} between the explanation text and system's corresponding recommendation.

To demonstrate the importance of sentiment alignment, we compare example output from two explainable recommendation algorithms (one proposed in this work, and another from \cite{li2017neural}) in Table \ref{tab:example}. Both algorithms strongly recommended restaurant A over B, as suggested by the corresponding large margins in their recommendation scores. Note such scores are not presented to the users in practice; even presented, they do not carry any detail about why an item is preferred by the algorithm. With Algorithm 1's explanations, one can easily recognize restaurant A is recommended because of better quality in its food and service. But on the contrary, it is much harder to comprehend the recommendations based on the Algorithm 2's explanations, as the presented difference become subtle, though their readability is comparable to Algorithm 1's. 
Two major reasons cost misaligned explanations in the second algorithm: 1) at training time, it only uses text reconstruction loss for explanation learning; 2) at inference time, the explanation is generated largely independently from the recommendation (as it only uses the predicted rating as an initial input for text generation).
The failure to align sentiment conveyed in the explanation text with the recommendations not only cannot help users make informed decisions, but also makes them confused or even doubt about recommendations, which is totally against the purpose of explainable recommendation.

We propose to enforce sentiment alignment in \emph{both} training and inference time for improved explainable recommendation. 
In particular, the learning of recommendation is modeled as a neural collaborative filtering problem \cite{he2017neural}, and the learning of explanation is modeled as a neural text generation problem \cite{sutskever2011generating}. We force the recommendation module to directly influence the learning of explanations by two means. First, we introduce two gated networks to our neural language model to fuse the intermediate output from the recommendation module to affect the word choice at every position of an explanation. Use examples shown in Table \ref{tab:example} again: given the currently generated content, the explanation module should properly choose the attribute words and corresponding sentiment modifiers (e.g., adjectives) to make their conveyed sentiment consistent with the recommendation module's prediction on this user-item pair. Second, a stand-alone sentiment regressor is added in between the two modules' output, such that its predicted sentiment score on the explanation text should be close to the given recommendation score. When discrepancy occurs, the explanation module is pushed to minimize the difference.
At inference time, all our treatments for sentiment alignment are kept. But since the explanation module has been learnt, the sentiment score gap is minimized by solving a constrained decoding problem. Because the sentiment regressor can only be applied to a complete text sequence, we use the Monte Carlo Tree Search algorithm \cite{kocsis2006bandit} for decoding with efficiency. Enforcing the alignment at inference time is vital, as it avoids the issue of decoupled output in existing explainable recommendation solutions.

We evaluate the proposed solution on both recommendation and explanation tasks, with particular focuses on the text quality, attribute personalization, and sentiment alignment of the generated explanations. The experiments are performed on Yelp and Ratebeer datasets in comparison with a rich set of popular baseline solutions. Empirical results show that our solution improves the performance on both tasks, with particularly improved explanation quality via its enhanced sentiment alignment. We also have our solution scrutinized under extensive user studies against competitive baselines. Positive user feedback suggests our explanations greatly help users gain a clearer understanding of the recommendations and make more accurate decisions.

\section{Related Work}

User-provided reviews have been popularly used as a proxy of explanations in explainable recommendations \cite{chen2018neural, wang2018reinforcement, li2019capsule}. One typical type of solutions directly extract representative text segments from existing reviews 
as explanations. For example, NARRE \cite{chen2018neural} uses attention to aggregate reviews to represent users and items for recommendation, in order to choose the most attentive reviews as explanations for each particular item. CARP \cite{li2019capsule} adopts the capsule network instead of attention for the same purpose. Wang et al. \cite{wang2018reinforcement} extend the idea with reinforcement learning to extract the most relevant review text segments that match a given recommender system's rating prediction. However, such explanations are limited to an item's existing reviews, some of which may not even be qualified as explanations (e.g., describing a personal experience). Moreover, these models only focus on selecting reviews to identify the items' characteristics, instead of addressing the reasons for a particular recommendation provided by the system. 
The lack of relevance hurts users' trust on both system-provided explanations and recommendations, and thus undermines the value of explainable recommendation. 

\begin{figure*}[th]
    \centering
    \includegraphics[width=0.9\linewidth]{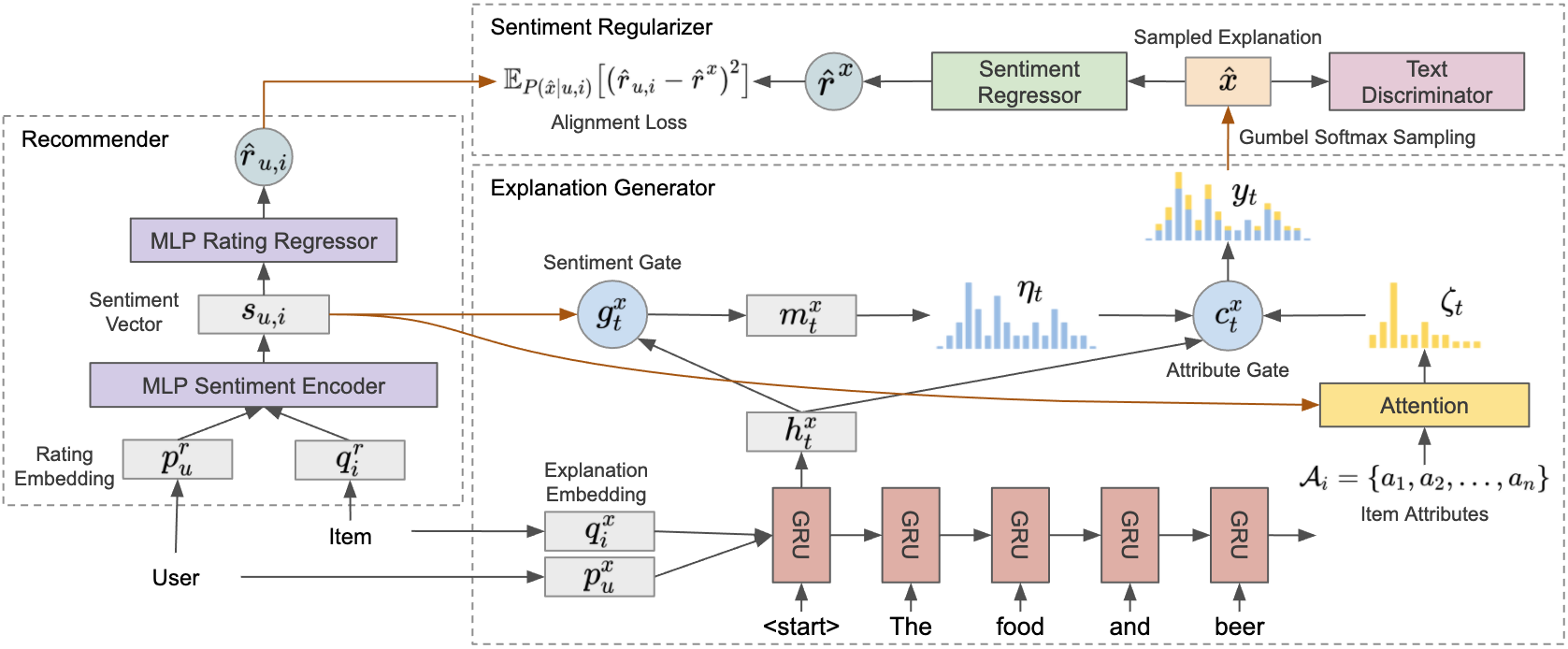}
    \vspace{-2mm}
    \caption{Model architecture of SAER. Sentiment alignment is explicitly enforced through three channels. First, SAER uses a shared sentiment vector to connect the recommender and explanation generator by the sentiment gate and attribute gate. Second, the sentiment regularizer samples generated explanations with Gumbel softmax and requires their carried sentiment (calculated by a pre-trained sentiment regressor) to match with the recommender's output score. Third, at inference time, constrained decoding is performed to ensure the alignment in the generated explanation. SAER also uses adversarial training to improve the explanations' readability in its sentiment regularizer.}
    \Description{The image contains three blocks: rater, generator and supervisor.}
    \label{fig:model_design}
    \vspace{-3mm}
\end{figure*}

Another family of solutions learn to generate explanations from reviews. Many of them learn to predict informative elements retrieved from reviews as explanations \cite{wang2018explainable, tao2019the, he2015trirank, ai2018learning}. As a typical example, MTER \cite{wang2018explainable} predicts items' attribute words and corresponding users' opinion words alone with its recommendations. Its explanations are generated by placing the predicted words into predefined templates, which however lack necessary expressiveness and diversity of nature language. Such robotic style explanations are usually considered less appreciated by users. To address this deficiency, neural language models have been applied to synthesize natural language explanations \cite{li2019persona, li2017neural, ni2019justifying, truong2019multimodal}. For example, NRT \cite{li2017neural} models explanation generation and item recommendation with a shared user-item embedding space, where its predicted recommendation rating is used as part of the initial state for corresponding explanation generation. MRG \cite{truong2019multimodal} integrates multiple modalities from user reviews, including ratings, text, and associated images, for explanation modeling, by treating them as parallel learning tasks. Neither the template-based or generation-based solutions paid enough attention to the sentiment alignment issue between recommendations and explanations. Although they jointly model recommendation and explanation (e.g., sharing embeddings), the objectives of training each module are still isolated. DualPC \cite{sun2020dual} realizes the importance of consistency between the two learning tasks, and introduces a duality regularization based on the joint probability of explanations and recommendations. 
However, the correlation imposed by duality does not have any explicit semantic meaning to the end users. In contrast, we require the output of models to be consistent in their carried sentiment, which is perceivable by an end user. 
Moreover, due to the required duality, DualPC has to use an ad-hoc approximation to break the coupling between the two models' output at inference time, which unfortunately hurts the learnt consistence between the two models. 
Our solution treats explanation as a dependent of recommendation, and solves a constrained decoding problem to infer the most aligned explanation at testing time accordingly.

\section{Sentiment Aligned Explainable Recommendation}

The problem of explainable recommendation can be formulated as follows: for a given pair of user $u$ and item $i$, the model outputs a personalized recommendation based on its computed score $r_{u,i}$ and a word sequence $x_{u,i}=\{w_1, w_2, \dots, w_n\}$ as its explanation. To learn such a model, we assume an existing training dataset, which includes a set of users $\cU$, items $\cI$, ratings $\cR$, attributes $\cA$, and explanation text $\cX$, denoted as as $\{\cU, \cI, \cR, \cA, \cX\}$. The attributes and explanations can be prepared from user-provided review corpora;
and we will introduce the procedure we adopted for this purpose later in the experiment section. We also define a vocabulary set $\cV = \{w_1, w_2, ..., w_{|\cV|}\}$ for explanation generation. We define attributes as items' popular properties mentioned in the review text, and thus they are a subset of vocabulary $\cA \subset \cV$.

Our model architecture for addressing explainable recommendation is shown in Figure \ref{fig:model_design}. It consists of three major components: 1) recommender, which takes a user and item pair $(u, i)$ as input to predict a recommendation score $\hat r_{u,i}$, which measures the affinity between $u$ and $i$; 2) explanation generator, which takes the $(u, i)$ pair as input and generates a word sequence $\hat x_{u,i}=\{w_1, w_2, \dots, w_n\}$ as the corresponding explanation; and 3) sentiment regularizer, which measures sentiment alignment between the generated explanation and recommendation. All three components closely interact with each other at \emph{both} training and inference time for improved explanation generation, especially for enhanced sentiment alignment. We name our solution Sentiment Aligned Explainable Recommendation, or SAER in short. 
Next, we will zoom into each component to introduce its design principle and technical details. 

\subsection{Personalized Recommendation}

As our focus in this work is not to design yet another recommendation algorithm, we adopted a latest neural collaborative filtering solution for the purpose \cite{he2017neural}. Arguably any latent factor models that explicitly learn user and item representations \cite{koren2009matrix,zhang2014explicit,wang2018explainable} can be adopted. In this section, we will only cover the most important technical details of our recommender's design, and leave interested readers to its original paper for more details.

We stack two Multi-Layer Perceptron (MLP) networks to predict the recommendation score $\hat r_{u,i}$ for a given $(u, i)$ pair. The first MLP encodes the $(u, i)$ pair to a latent sentiment vector $\bs_{u,i}\in\mathbb{R}^{d^r_s}$, and the second MLP maps the sentiment vector $\bs_{u,i}$ into the numerical rating $\hat r_{u,i}$. We refer to the first MLP as \emph{sentiment encoder} and the second one as \emph{rating regressor}. Instead of using the predicted score $\hat r_{u,i}$ to influence explanation generation, we choose to inform the explanation generator by the encoded sentiment vector $\bs_{u,i}$. We defer the details of this design to the next section.  

In the recommendation module, we define the latent embedding matrices for users and items as $P^r \in \mathbb{R}^{d^r \times |\cU|}$ and $Q^r \in \mathbb{R}^{d^r \times |\cI|}$ respectively, where $d^r$ is the dimension of the embedding vectors. 
The sentiment encoder concatenates the embedding vector $p^r_u$ and $q^r_i$ as its input and passes it through multiple layers with leaky ReLU activation to get the sentiment vector $\bs_{u,i}$ encoded.
Besides its use in the explanation generator, $\bs_{u,i}$ is then mapped by the rating regressor through another set of multi-layer leaky ReLUs to get the final recommendation score ${\hat r}_{u,i}$. 

In addition to the popularly used Minimal Squared Error (MSE) \cite{chen2018neural, li2017neural} to train our recommender, we also introduce a pairwise hinge loss to improve the trained recommender's ranking performance. Specifically, for each user $u$, we collect a set of personalized item pairs $\cB_u = \{(i, j)|r_{u,i} > r_{u,j}\}$, where $i$ and $j$ are two items rated by user $u$ and one is preferred than another as observed in the training dataset. We did not use the popular BPR loss \cite{rendle2012bpr}, because it tends to push ratings to extreme values, which is inconsistent with our sentiment regularizer's requirement to be explained later. 

Based on the rating set $\cR$ and personalized item pair set $\{\cB_u\}_{u\in\cU}$, the loss for recommender training is defined as:
\begin{equation*}
    L^r = \frac{1}{|\cR|}\sum_{r_{u,i} \in \cR} \big(\hat{r}_{u,i} - r_{u,i}\big)^2 +  \sum_{u\in\cU}\frac{\lambda_h}{|\cB_u|}\sum_{(i,j) \in \cB_u}\max\big(0, \beta - (\hat{r}_{u,i} - \hat{r}_{u,j})\big)
\end{equation*}
where $\beta > 0$ is a hyper-parameter to control the separation margin, i.e., it penalizes the model when the predicted difference between $\hat r_{u,i}$ and $\hat r_{u,j}$ is smaller than $\beta$, and $\lambda_h$ is the coefficient to control the balance between MSE loss and pairwise hinge loss. 

\subsection{Explanation Generation}
Motivated by the success of neural language generation, we appeal to a Recurrent Neural Network (RNN) model with Gated Recurrent Units (GRUs) \cite{chung2014empirical} for explanation generation. 
To make the generation related to the user and item, we first map the input user $u$ and item $i$ to their embeddings $p^x_u$ and $q^x_i$ with the latent matrices $P^x \in \mathbb{R}^{d^x \times |\cU|}$ and $Q^x \in \mathbb{R}^{d^x \times |\cI|}$ learnt by the explanation generator. 
We should note this set of embeddings are different from those used in the recommender (i.e., $P^r$ and $Q^r$), as they should characterize different semantic aspects of users and items (ratings vs., text). 
We hence use superscript $x$ to indicate variables and parameters related to explanation generator.
To generate explanation text, the embeddings are concatenated and linearly converted into the initial RNN hidden state; and then the GRU generates hidden state $\bh^x_t\in\mathbb{R}^{d^x_h}$ at position $t$ with previous state $\bh^x_{t-1}$ and input word $w_t$, and predicts the next word $w_{t+1}$ recursively.

We initialize RNN with pretrained GloVe word embeddings $V \in \mathbb{R}^{d^x_\mathbf{v} \times |\cV|}$ \cite{pennington2014glove}.

Though similar model design has been used for explanation generation \cite{li2017neural,sun2020dual}, this straightforward application of RNN can hardly generate satisfactory explanations, where two issues are left open. First, a good explanation is expected to be personalized and specific about the recommendation; generic content, such as ``\emph{this is a nice restaurant}'' can never be informative. It is important to explain the recommended item by the user's most concerned attributes \cite{wang2010latent}. Second, the sentiment carried in the explanation, especially on the mentioned attributes, should be explicit and consistent with the recommendation (as shown in our case study in Table \ref{tab:example}). There is no guarantee that a simple RNN can satisfy both requirements.

We enhance our generator design with two gated sub-networks upon GRU to address the aforementioned issues.
First, we design a sub-network, named attribute gate, to guide attribute word generation with respect to the input user-item pair and the predicted recommendation sentiment. The attribute gate is built based on a pointer network (or copy mechanism) \cite{see2017get, zeng2016efficient}, which decides whether the current position should mention an attribute word and the corresponding distribution of attribute words based on the generation context. 
To make the choice of attribute word specific to the item, for each item $i$ we build an attribute set with all attribute words that appear in $i$'s associated training explanation text: $\cA_i = \big\{a_k|a_k \in \{x_{u,i} | u \in \cU\} \big\}$. 
To make the attribute choice depend on the already generated content, we attend on the concatenation of the current position's RNN hidden state $\bh^x_t$ and sentiment vector $\bs_{u,i}$ to compute the distribution of these attribute words,
\begin{equation}
\mathbf{z}_{t,k} = [\bh^x_t, \bs_{u,i}]^\top W^x_z \mathbf{v}_{a_k},  \forall k, a_k \in\cA_i;
    ~~ \mathbf{\zeta}_t = softmax(z_t), 
\label{eq_attribute_gate}
\end{equation}
where $W^x_z \in \mathbb{R}^{(d^x_h + d^r_s) \times d^x_\mathbf{v}}$ and $\mathbf{v}_{a_k}$ is the word embedding of attribute $a_k$. $\mathbf{z}_{t,k}$ is computed for every $a_k$ in $\cA_i$, i.e., $\mathbf{z}_t = \{z_{t,1}, z_{t,2} \\ ... z_{t, |\cA_i|}\}$. $\mathbf{\zeta}_t$ is the resulting attribute word distribution at position $t$. 
For better performance, an extra linear transformation can be applied to $\bh^x_t$ to compress it into a lower dimension before computing attention, which helps avoid overfitting attentions to the text generation context but ignoring the sentiment context.      

To decide if we need to generate an attribute word using Eq \eqref{eq_attribute_gate} at position $t$, we compute the copy probability with respect to the current context $\bh^x_t$ by $c^x_t = \sigma(W^x_c \bh^x_t + b^x_c)$, where $\sigma(\cdot)$ is the sigmoid function, $W^x_c \in \mathbb{R}^{d^x_h}$ and $b^x_c \in \mathbb{R}$. $c^x_t$ allows us to mix the vocabulary distribution predicted by GRU and attribute word choice to get our final word distribution at position $t$.

Second, we design a sentiment gate to fuse the sentiment vector $\bs_{u,i}$ to align sentiment in the generated explanation text. Our key insight is that not all words convey sentiment, we need to choose the right word at the right place to express consistent sentiment as needed by the recommender. 
Similar to our attribute gate design, we apply a soft gate 
to decide how each position is related to the intended sentiment. At position $t$, the sentiment gate calculates a ratio $g^x_t$ with respect to the RNN's hidden state $\bh^x_t$. The sentiment vector $\bs_{u,i}$ is then weighted and merged with $\bh^x_t$,
\begin{equation}
    g^x_t = \sigma(W^x_g \bh^x_t + b^x_g), ~~\mathbf{m}^x_t = tanh\big(\bh^x_t + g^x_t (W^x_m \bs_{u,i} + b^x_m)\big)
    \label{eq_sentiment_gate}
\end{equation}
where $W^x_g \in \mathbb{R}^{d^x_h}$ and $b^x_g \in \mathbb{R}$ produce a scalar $g^x_t$. $\mathbf{m}^x_t$ is the sentiment fused latent vector to predict the vocabulary distribution for position $t$.  
Because not all words are about sentiment, to better differentiate the positions where the intended sentiment needs to be expressed from the rest, we impose sparsity on the learned gate value $g^x_t$ using L1 regularization at training time. In other words, the gate is open only when necessary.

We compute the final word distribution by consolidating the outputs of the two gated sub-networks (Eq \eqref{eq_attribute_gate} and \eqref{eq_sentiment_gate}). First, the sentiment fused latent vector $\mathbf{m}^x_t$ is fed through a linear layer to calculate the vocabulary distribution $\mathbf{\eta}_t = softmax(W^x_\mathbf{v} \mathbf{m}^x_t + b^x_\mathbf{v})$, where $W^x_\mathbf{v} \in \mathbb{R}^{|\cV| \times d^x_h}$ and $b^x_\mathbf{v} \in \mathbb{R}^{|\cV|}$. Second, the vocabulary distribution $\mathbf{\eta}_t$ and attribute word distribution $\mathbf{\zeta}_t$ are merged to obtain the final word distribution with respect to the copy probability $c^x_t$, i.e., $\mathbf{y}_t = (1 - c^x_t)  \mathbf{\eta}_t + c^x_t \mathbf{\zeta}_t$, where the value of $w_k$ in $\mathbf{\zeta}_t$ is $0$ if $w_k$ is not an attribute word.

The objective for explanation generation is to minimize the negative log-likelihood loss (NLL) on the training explanation set $\cX$,
\begin{equation*}
L^x = - \sum_{x \in \cX}\sum_{w_t \in x}\log{\mathbf{y}_{t}(w_t)} + \lambda_g\sum_{x \in X}\sum_{w_t \in x}|g^x_t|
\end{equation*}
where $\mathbf{y}_{t}(w_t)$ is the resulting probability of word $w_t$ and $\lambda_g$ is the coefficient for the L1 regularization of the sentiment gate values.

\subsection{Sentiment Alignment}

Though our sentiment gate design (Eq \eqref{eq_sentiment_gate}) introduces predicted sentiment from the recommender to the explanation generator, it is still insufficient to guarantee sentiment alignment, for three major reasons. First, word-based NLL training cannot maintain the whole sentence's sentiment. For example, as the number of sentiment words in an explanation is less than the number of non-sentiment words, the training is affected more by those non-sentiment words. This weakens its prediction quality on sentiment words. Second, the explanation generator might utilize the sentiment vector differently as the recommender does, so that the recommendation rating might diverge from the sentiment carried by the explanation. Third, the generation process at the inference stage works differently from the training stage \cite{ranzato2015sequence}: at inference time, the previously decoded word is used as the input for the next word prediction, instead of the ground-truth word as at the training time. Hence, the learnt text pattern might not be fully exploited at the inference time. 

We introduce the sentiment regularizer to close the loop between the recommender and explanation generator. It uses a stand-alone sentiment regressor to predict the sentiment rating $\hat r^x$ on the generated explanation text $\hat x_{u,i}$ for user-item pair $(u,i)$, and requires the explanation generator to match the rating $\hat r_{u,i}$ from the recommender accordingly. We do not have any particular assumption about the sentiment regressor; and any state-of-the-art regression models can be leveraged \cite{DBLP:journals/ftir/PangL07}. 
In this work, we employed an MLP on top of a bidirectional RNN text encoder with inner attention for rating regression, and denote it as $f^R(x)\to r^x$. We pre-train this regressor based on ground-truth $\{\cR,\cX\}$ in the training set; and fix the learnt model thereafter.

To enforce sentiment alignment by the predicted ratings, we introduce a new loss to the training of our explanation generator,
\begin{equation}
\label{cr_loss}
L^{a} = \sum_{u\in\cU,i\in\cI}\mathbb{E}_{P(\hat{x}|u,i)}\big[(\hat{r}_{u,i} - f^R(\hat{x}))^2\big]
\end{equation}
where $P(\hat{x}|u,i)$ is the probability of generating $\hat{x}$ for the given $u$ and $i$. We should note this loss is not necessarily restricted to the observed $(u,i)$ pairs in the training set; instead, it could be any pairs of them, since both the recommender and explanation generator can generate output on any given $(u,i)$ pair. It thus enables data augmentation for sentiment alignment.  

However, because the word distribution is categorical, the generation of $\hat{x}$ is not differentiable. It makes direct optimization with respect to Eq \eqref{cr_loss} infeasible. As a result, we appeal to Gumbel softmax \cite{jang2016categorical} to obtain approximated  gradient of sampling from a categorical distribution. 
Briefly, Gumbel softmax reparameterizes the randomness in sampling by a gumbel distribution and simulates a relaxed one-hot vector with softmax. As we need a strict one-hot vector to represent each single word, we adopt the Straight-Through (ST) Gumbel softmax estimator \cite{jang2016categorical}. For each $(u,i)$ pair in Eq \eqref{cr_loss}, we back-propagate the gradient from $L^a$ to the explanation generator to improve the quality of sentiment alignment on the whole sequence.

Unfortunately, this new sentiment alignment loss might also attract the generation process
to produce unreadable sequences, which however match the intended sentiment ratings.
For example, the sentiment regressor may give a very positive rating to an unnatural sentence ``\textit{good good good good}'', when the recommender also happens to predict a high rating for this item. Giving a higher weight to the NLL loss $L^x$ in explanation learning cannot address this issue, as it cannot inform why a particular sequence should not be generated.

To improve the readability of our generated explanation, we introduce a 
text discriminator $f^D$, which learns to differentiate the authentic explanations from the generated ones, to guide the explanation generation as well. Our design allows any text classifier. In this work, we used an MLP binary classifier on top of a bidirectional RNN encoder for the purpose. We train the discriminator using cross-entropy loss with the ground-truth explanations $x$ as positive and the generated explanations $\hat{x}$ as negative,
\begin{equation*}
    L^D = - \frac{1}{|\cX|} \sum_{x \in \cX} \log f^D(x) - \sum_{u\in\cU, i\in\cI}\mathbb{E}_{P(\hat{x}|u,i)}\big[\log(1 - f^D(\hat{x}))\big]
\end{equation*}
Correspondingly, another objective of explanation generation is to fool the discriminator, i.e., the adversarial loss,
\begin{equation*}
    L^{c} = - \sum_{u\in\cU, i\in\cI}\mathbb{E}_{P(\hat{x}|u,i)}\big[ \log f^D(\hat{x})\big]
\end{equation*}
This loss also requires sampled explanations $\hat{x}$ as the input, like the alignment loss defined in Eq \eqref{cr_loss}. The same Gumbel softmax sampling technique is used for end-to-end training.

As we pointed out before, addressing the sentiment alignment issue in training alone is still insufficient, we introduce a constraint-driven decoding strategy to enhance the alignment at the inference stage as well. 
Similarly as in training, we use MSE to quantify the difference between the rating predicted from the explanation text and that from the recommender. But at the inference stage, since the explanation generator has been trained and fixed, the discrepancy can only be minimized by varying the generation of explanation text, e.g., trial and error.

Because the sentiment regressor can only be applied to a complete sequence, the search space is too large to enumerate by the generator. 
Hence, we treat generating explanation $\hat x$ at inference time as a sequence of decision making, where each action is to generate a word $w_t$ at position $t$, given its already generated prefix as state. But we do not have feedback on the actions, until we complete $\hat x$; and the return for taking the series of actions can be measured by $Q(\hat{x}; \hat{r}_{u,i}) = [\hat{r}_{u,i} - f^R(\hat{x})]^2$. 
To find a policy that minimizes return (since we want to reduce the discrepancy), we need to estimate the value function under each state. This is a well studied problem in reinforcement learning, and it can be effectively addressed by Monte Carlo Tree Search (MCTS) \cite{kocsis2006bandit}. Basically, we estimate the value function using our trained explanation generator for roll-out. When at position $t$ for generating $\hat x_{u,i}$, we will sample $n$ complete sequences for every action $w$ using the current prefix $\{w_1, w_2,\dots,w_{t-1}\}$, following the distribution specified by the explanation generator: $\hat{X}_{u, i, t}(w) = \big\{\hat{x}_k = MCTS_{u,i}(w_1, w_2,\dots,w_{t-1}, w)\big\}^n_{k=1}$. Then the value of taking action $w$ at position $t$ can be estimated by,
\begin{equation*}
Q(w_1, w_2,\dots,w_{t-1}, w; \hat{r}_{u,i}) = \frac{1}{|\hat{X}_{u,i,t}(w)|} \sum_{\hat{x}_k \in \hat{X}_{u,i,t}(w)} Q(\hat{x}_k, \hat{r}_{u,i})
\end{equation*}
Based on the estimated values, we can take the action that minimizes the value. A recent study \cite{holtzman2019curious} suggests that top-k sampling oftentimes avoids bland and repetitive content compared to more commonly used greedy decoding strategies, such as top-1 or beam search.
Therefore, we integrate our MCTS with top-k sampling, i.e., at each decoding position $t$, we sample $k$ most likely words according to word distribution $\mathbf{y_t}$ and then use MCTS to select the one that minimizes the estimated value under given state.

A vanilla implementation of MCTS is expected to be expensive and slow in our problem, as it needs to complete the sequence at each position from an RNN model for multiple times. Fortunately, our sentiment gate design provides a short path for efficient sampling: as sentiment is only carried by a small number of words, there is no need to conduct such expensive sampling procedure at every position. Instead, we only need to perform MCTS at positions where sentiment is expressed. Hence, we set a threshold on the sentiment gate's value to decide when to perform MCTS. When the gate's value is below the threshold, we will directly sample from the top-k words of the explanation generator's prediction.

\subsection{End-to-End Model Training}

Putting together the three components in our proposed explainable recommendation solution SAER, the overall objective of our model training is formulated as:
\begin{equation*}
    J = \min_{\Theta}\big(\lambda_r L^r + \lambda_x L^x + \lambda_{a} L^{a}  + \lambda_{c} L^{c} + \lambda_n ||\Theta||^2\big)
\end{equation*}
where $\Theta$ is the complete set of model parameters, and $\{\lambda_r, \lambda_x, \lambda_a, \lambda_c\}$ are the corresponding coefficients to control the relative importance of each component in model training. We also include an L2 regularization for the model parameters $\Theta$, weighted by its coefficient $\lambda_n$. The parameters are then effectively estimated end-to-end with stochastic gradient optimizer of Adam \cite{kingma2014adam}.

However, due to our model's complex structure, it is challenging to fully unleash the optimizer's potential on its own. Therefore, we split the whole training process into five stages. First, estimate the sentiment regressor on $\{\cX, \cR\}$, as it does not depend on the other parts of our model. Second, pre-train the recommender on $\{\cU, \cI, \cR\}$ till convergent. This step is essential to learn a good sentiment encoder whose output will be used to inform the explanation generator. Third, freeze the recommender and train the generator on $\{\cU, \cI, \cA, \cX\}$ with negative log-likelihood loss only. 
We found in our experiments that generation learning was more difficult than recommendation learning. First training the explanation generator separately can help align the training of both modules later. Fourth, after the separate training converges, start joint training of the recommender and explanation generator. This step allows the model to align the sentiment representation from both modules. At last, freeze the recommender, and turn on the sentiment regularizer to further improve the explanation generator. At this stage, the explanation discriminator and generator are trained in turn.

\section{Experimental Evaluation}
We quantitatively evaluate our model's performance on personalized recommendation and explanation generation in two different domains: restaurant recommendation on Yelp reviews \footnote{https://www.yelp.com/dataset} and beer recommendation on Ratebeer reviews \cite{julian2012learning}. Our model is compared against a set of state-of-the-art baselines on both offline data and user studies, where encouraging improvements in both recommendation and explanation tasks are obtained.

\subsection{Experiment Setup}
\subsubsection{Data Pre-Processing}

As the attributes are not directly provided in these two review datasets, we use the Sentires toolkit \cite{zhang2014users} to extract attribute words from reviews and manually filter out inappropriate ones based on domain knowledge.
Although reviews are directly treated as explanations in many previous studies \cite{chen2018neural, wang2018reinforcement}, a recent work \cite{ni2019justifying} suggests a large portion of review content is only about subjective emotion and thus does not qualify as explanations, e.g., ``\textit{I love the food}''. An informative explanation should depict the details of items, e.g., their attributes, to help users perceive the exact reason behind recommendations, e.g., ``\textit{the fish is fresh}''. Therefore, we restrict ourselves to sentences containing attribute words as explanations in our experiments.

On top of the crafted explanations, we select 20,000 most frequent words and map others to unknown to build the vocabulary.
Finally, as lots of users and items only have very few reviews in the datasets, we apply recursive filtering as in \cite{wang2018explainable} to refine the datasets and alleviate this sparsity issue. 
The resulting statistics of the datasets are summarized in Table \ref{tab:stats}. 

\begin{table}[t]
    \caption{Statistics of the processed datasets.}
    \vspace{-2mm}
    \label{tab:stats}
    \begin{tabular}{|c|cccc|}
    \hline
    Dataset  & \# Users & \# Items & \# Reviews & \# Attributes \\ \hline
    Yelp     & 15,642   & 21,525   & 1,108,971  & 498           \\
    Ratebeer & 3,895    & 6,993    & 1,073,762  & 333           \\ \hline
    \end{tabular}
    \vspace{-4mm}
\end{table}

\subsubsection{Baselines}
To evaluate the personalized recommendation performance, we used the following recommendation baselines:
\begin{itemize}[leftmargin=*]
  \item[-] \textbf{NMF}: Non-negative Matrix Factorization \cite{lee2001algorithms}. A widely used latent factor model, which decomposes the rating matrix into lower dimensional matrices with non-negative factors. 
  \item[-] \textbf{SVD}: Singular Value Decomposition \cite{koren2008factorization}. It utilizes rating matrix as input for learning user and item representations.
  \item[-] \textbf{NCF}: Neural Collaborative Filtering \cite{he2017neural}. It is a modified matrix factorization solution which adopts neural networks to model the nonlinear vector operations. 
\end{itemize}

We also include two explainable recommendation baselines that can output natural language sentences as explanations for comparing both the recommendation and explanation quality:
\begin{itemize}[leftmargin=*]
  \item[-] \textbf{NARRE}: Neural Attentional Regression model with Review-level Explanations \cite{chen2018neural}. It learns the usefulness of the existing reviews through attention,   and incorporates the review to enrich user and item representations for rating prediction. 
  To fit in our evaluation, we select sentences from its most attentive reviews as explanations. 
  \item[-] \textbf{NRT}: Neural Rating and Tips Generation \cite{li2017neural}. A multi-task learning solution for rating regression and content generation. It uses the predicted recommendation score to create initial states for content generation. 
\end{itemize}

\begin{table*}[t]
    \caption{Evaluation of personalized recommendation in terms of rating prediction (RMSE, MAE) and item ranking (NDCG).}
    \vspace{-2mm}
    \label{tab:recomm_eval}
    \begin{tabular}{|c|ccccc|ccccc|}
        \hline
         &
        \multicolumn{5}{c|}{Yelp} &
        \multicolumn{5}{c|}{Ratebeer} \\
        Model & RMSE & MAE & NDCG@3 & NDCG@5 & NDCG@10 & RMSE & MAE & NDCG@3 & NDCG@5 & NDCG@10 \\
        \hline
        NMF & 1.1034 & 0.8164 & 0.3777 & 0.5067 & 0.7344 & 2.2228 & 1.6609 & 0.5143 & 0.6334 & 0.7766 \\
        SVD & 1.0286 & 0.7975 & 0.3924 & 0.5246 & 0.7519 & 2.2942 & 1.6474 & 0.4952 & 0.6120 & 0.7593 \\
        NCF & 1.0532 & 0.8251 & 0.3850 & 0.5150 & 0.7420 & 2.0857 & 1.5002 & 0.5421 & 0.6621 & 0.8004 \\
        NARRE & 1.0275 & 0.8035 & 0.3918 & 0.5230 & 0.7509 & 2.0714 & 1.4975 & 0.5464 & 0.6641 & 0.8030 \\
        NRT & 1.0254 & 0.8017 & 0.3947 & 0.5262 & 0.7540 & 2.0743 & 1.4922 & 0.5436 & 0.6620 & 0.8008 \\
        SAER & \textbf{1.0190} & \textbf{0.7948} & \textbf{0.3953} & \textbf{0.5278} & \textbf{0.7553} & \textbf{2.0628} & \textbf{1.4842} & \textbf{0.5468} & \textbf{0.6648} & \textbf{0.8034} \\
        \hline
    \end{tabular}
    \vspace{-2mm}
\end{table*}

\subsection{Quality of Personalized Recommendations}
We evaluate the recommendation quality both in terms of rating prediction (by RMSE and MAE) and item ranking performance (by NDCG@\{3,5,10\} \cite{jarvelin2017ir}). The results are shown in Table \ref{tab:recomm_eval}.
SAER demonstrates better performance in all metrics on both datasets. 
In particular, thanks to the introduced hinge loss for pairwise ranking, SAER demonstrates improved ranking performance against all baselines, which only modeled recommendation as a rating prediction task. 
The performance difference among NCF, NRT and SAER is worth noting. Although their rating prediction modules all use MLP, NRT and SAER additionally leverage the content information for improved recommendation quality. Improvements from SAER against NARRE and NRT demonstrate that our sentiment vector and corresponding soft gate design better distill and exploit review data for joint learning. Again, as our focus in this work is not on improving recommendation quality, but more on explanation. Next, we will dive into our extensive evaluations about the generated explanation.

\subsection{Quality of Generated Explanations}
We evaluate the quality of our generated explanations from three perspectives: text quality, attribute personalization, and sentiment alignment.
We introduce two variants of our model to better analyze the effects of our sentiment regularizer and constrained decoding strategy. 
1) SAER (topk), it removes sentiment regularization and decodes by top-k sampling, such that sentiment alignment is only introduced by the soft gates, without the alignment loss, nor the constrained decoding; 2) SAER (reg + topk), it uses sentiment regularization (i.e., the alignment loss) and decodes by top-k sampling, such that sentiment alignment is only enforced at training time.

\begin{table}
    \caption{BLEU scores of generated explanations.}
    \vspace{-3mm}
    \label{tab:bleu}
    \begin{tabular}{|cc|ccc|}
        \hline
        
        Dataset & Model & BLEU-1 & BLEU-2 & BLEU-4 \\
        \hline
        & NARRE & 20.46 & 5.72 & 2.12 \\
        & NRT & 26.25 & 8.84 & 2.97 \\
        Yelp & SAER (topk) & 27.43 & 9.53 & 3.18 \\
        & SAER (reg + topk) & 28.69 & 10.29 & 3.37 \\
        & SAER & \textbf{28.88} & \textbf{10.44} & \textbf{3.44} \\
        \hline
        & NARRE & 29.78 & 9.47 & 3.27 \\
        & NRT & 42.16 & 17.54 & 5.63 \\
        Ratebeer & SAER (topk) & 43.92 & 19.60 & 6.56 \\
        & SAER (reg + topk) & 45.69 & 21.09 & 7.02 \\
        & SAER & \textbf{46.01} & \textbf{21.60} & \textbf{7.32} \\
        \hline
    \end{tabular}
    \vspace{-2mm}
\end{table}

\subsubsection{Quality of Generated Text.}
We measure the quality of generated explanation text with BLEU \cite{papineni2002bleu},
and report the results in Table \ref{tab:bleu}. The extraction-based NARRE performed clearly worse than other generation-based models. This is because the synthesized natural language explanations are not limited to the existing review content and is more flexible to customize for a particular user-item pair. NRT uses the predicted ratings in the initial state for content generation, in comparison to the sentiment vectors used in SAER. The performance gap between NRT and SAER (topk) suggests that our sentiment vectors are more expressive and the two soft gates can better guide explanation generation throughout the process, than only affecting RNN's initial state. The additional gain brought by the sentiment regularizer in SAER (reg + topk) and constrained decoding in SAER  highlights the benefits of sentiment alignment in both training and inference time.

\begin{table}
    \caption{Performance of attribute prediction in generated explanations.}
    \vspace{-3mm}
    \label{tab:fp}
    \begin{tabular}{|c|cc|cc|}
    \hline
     & \multicolumn{2}{c|}{Yelp} & \multicolumn{2}{c|}{Ratebeer} \\
    Model & Precision & Recall & Precision & Recall \\
    \hline
    NARRE & 0.1415 & 0.1906 & 0.2176 & 0.2245 \\
    NRT & 0.1791 & 0.1997 & 0.3443 & 0.1720 \\
    SAER (topk) & 0.2024 & 0.2297 & 0.3523 & 0.2554 \\
    SAER (reg + topk) & 0.1992 & 0.2319 & 0.3549 & 0.2614 \\
    SAER & \textbf{0.2115} & \textbf{0.2391} & \textbf{0.3702} & \textbf{0.2677} \\
    \hline
    \end{tabular}
    \vspace{-2mm}
\end{table}

\subsubsection{Attribute Personalization}
An informative explanation should cover the users' most concerned aspects. We evaluate such performance in terms of attribute personalization. 
For each user-item pair, we evaluate precision and recall of attribute word in the algorithms' explanations against ground-truth explanations. 
The results in Table \ref{tab:fp} show the improvement brought by our attribute gate, which is proved to be effective in predicting users' most concerned attributes. As the two baselines do not pay attention to items' attributes when generating the explanations, their quality in providing attribute-level explanations is much worse.

\begin{table}
    \caption{Sentiment alignment evaluation of decoded explanations by RMSE. PD is the RMSE between explanation rating and predicted rating, and GT is the RMSE between explanation rating and ground-truth rating.}
    \vspace{-2mm}
    \label{table_sen}
    \begin{tabular}{|c|cc|cc|}
        \hline
          & \multicolumn{2}{c|}{Yelp} & \multicolumn{2}{c|}{Ratebeer} \\
          & PD & GT & PD & GT \\
        \hline
        NARRE & 1.0932 & 1.4950 & 2.0996 & 2.9641 \\
        NRT & 0.6676 & 1.2086 & 2.3302 & 3.1304 \\
        SAER (topk) & 0.6908 & 1.2216 & 2.1727 & 3.0026 \\
        SAER (reg + topk) & 0.6242 & 1.1849 & 1.6985 & 2.6769 \\
        SAER & \textbf{0.5505} & \textbf{1.1503} & \textbf{1.5911} & \textbf{2.6042} \\
        \hline
    \end{tabular}
    \vspace{-3mm}
\end{table}

\subsubsection{Sentiment Alignment Between Ratings and Explanations}
Offline evaluation of sentiment alignment is not easy, since it should be evaluated by the end users who receive the recommendation and explanation. In addition to depending on user studies to evaluate this aspect (reported in the next section), we also use our pre-trained sentiment regressor for an approximated offline evaluation. For a generated explanation, we infer its carried sentiment by our sentiment regressor. We then compute the RMSE between the inferred rating from explanation and that predicted by the recommendation module (marked as PD). This measures sentiment difference between the recommendation and corresponding explanation. We also compare the inferred rating against the ground-truth rating (marked as GT) as a reference. The results are presented in Table \ref{table_sen}. 
Without our sentiment regularizer, SAER (topk) can already significantly outperform the baselines on Yelp, which demonstrates the utility of our two gated network design for sentiment alignment. And the alignment loss and constrained decoding further push SAER's explanations closer to its recommendations. Compared to the ground-truth rating, sentiment carried by the explanation is closer to the recommender's prediction. We hypothesize that this can be caused by the difficulty to predict ground-truth rating: as reported in Table \ref{tab:recomm_eval}, the accuracy of the recommender's rating prediction is at around the same level.

\begin{table}[t]
    \caption{Agreement rate between the model's predicted item ranking and the users perceived ranking based on the provided explanations.}
    \vspace{-2mm}
    \label{tab:ar}

    \begin{tabular}{|c|c|cc|}
\hline
\multirow{3}{*}{\begin{tabular}[c]{@{}c@{}}Agreement\\ Rate\end{tabular}} & Model & gap>0.5 & gap$\leq$0.5 \\ \cline{2-4} 
                                                                           & NRT   & 64.76\%              & 52.62\%             \\
                                                                           & SAER  & \textbf{73.10}\%              & \textbf{61.90}\%             \\ \hline
\end{tabular}
    \vspace{-3mm}
\end{table}

\section{User Study}

We conduct extensive user studies on Amazon Mechanical Turk to evaluate our explanations' utility to real users. 
We chose the restaurant recommendation task based on the Yelp dataset, as it is more familiar by general users. 

We design two separate tasks. The first task focuses on evaluating if the generated explanations can help users make more informed decisions about the recommendations.  
In this task, we randomly pair items with different ratings predicted by a tested algorithm, and ask participants to read the corresponding explanations before choosing the item they perceived as the better one. We then evaluate the agreement rate between participants' choices and the algorithm's predictions. 
Specifically, without showing the actual predicted scores to participants, we present the corresponding explanations and require them to answer the following question:
\begin{itemize}[leftmargin=*]
\item[]  \textit{``After reading the provided explanations, which restaurant would you like to visit? You are expected to judge the quality of the recommended restaurant based on the provided explanations, and then choose the one with better quality.''}
\end{itemize}
In this experiment, we only adopted NRT as the baseline, because NARRE's explanations are item-based and thus not personalized for individual users.

To demonstrate the explanations' sentiment sensitivity towards recommendations, i.e., whether a user can correctly tell the difference between the two recommended items by reading the explanations, we group the results by the gap between the two items' predicted scores, and choose 0.5 as the threshold. We collected 420 responses for each model in each group, resulting 1,680 responses in total. The results are presented in Table \ref{tab:ar}. Both models' explanations are reasonably discriminative when the rating gap is larger. 
But it is more challenging to explain the difference when the recommendation scores are close. When the gap is smaller than $0.5$, the agreement rate on NRT's results dropped to around 50\%, which suggests users can barely perceive the differences by reading the explanations. In contrary, users can better tell the difference from SAER's explanations for making informed decisions.

\begin{table}
    \caption{Up-vote rate of explanations' helpfulness.}
    \vspace{-2mm}
    \label{tab:rh}
    \begin{tabular}{|c|c|cc|}
\hline
                                                                             & Model           & Positive         & Negative         \\ \hline
\multirow{4}{*}{\begin{tabular}[c]{@{}c@{}}Up-vote\\ Rate\end{tabular}}      & NARRE           & 23.33\%          & 42.86\%          \\
                                                                             & NRT             & 50.69\%          & 26.98\%          \\
                                                                             & GT              & 46.77\%          & \textbf{46.76\%} \\
                                                                             & SAER            & \textbf{57.58\%} & 41.76\%          \\ \hline
\multirow{3}{*}{\begin{tabular}[c]{@{}c@{}}Paired \\ t-test\end{tabular}} & SAER v.s. NARRE & 0                & 0.6786           \\
                                                                             & SAER v.s. NRT   & 0.0046           & 0                \\
                                                                             & SAER v.s. GT    & 0                & 0.9810           \\ \hline
\end{tabular}
    \vspace{-3mm}
\end{table}

The second task studies whether the explanations can help users comprehend the reason of a recommended item. In particular, we ask the participants to compare explanations of the same recommended item but provided by different algorithms, and then select the most useful ones. 
We categorize the items as recommended (top ranked items) or not recommended (bottom ranked items) to study if the model can provide the correct explanations for both categories. For each item, we shuffle the explanations from different models for participants to select from. To help participants better judge the explanation quality, we also provide the restaurant's name and cuisine type.
Specifically, we ask one of the following questions according to whether the item is recommended:
\begin{itemize}[leftmargin=*]
  \item[-] \textbf{Positive recommendation}: \textit{``Which of the following explanations help you the most to understand why you should pay attention to the recommended restaurant?''}
  \item[-] \textbf{Negative recommendation}:  \textit{``Which of the following explanations help you the most to understand why our system believes the restaurant is NOT a good fit for you?''}
\end{itemize}
We choose NARRE, NRT and ground-truth explanations for comparison; and compare them by their received helpfulness votes.

We collected 904 responses for positive  recommendations and 752 for negative. Table \ref{tab:rh} reports the up-vote rates of the explanations from different models and the results of paired t-test. 
In positive recommendations, the generation-based methods, i.e., SAER and NRT, are preferred; and SAER significantly outperforms others. This reveals that the common and concise syntax and vocabulary of synthesized language are preferred in the explainable recommendation scenario, because users can more easily understand the explanations. On the negative recommendations, however, the results are mixed. SAER is still preferred over NRT, but worse than NARRE and ground-truth. The key reason is the inherent data bias: the Yelp dataset contains much more positive reviews than negative ones. Such imbalance makes SAER reluctant to generate negative explanations and less trained for negative content. Hence, its generated explanations cannot strongly justify the negative recommendations. But from a different perspective, this result also echoes the importance of aligned sentiment in explainable recommendation.

\section{Conclusion and Future Work}

In this paper, we present a new explainable recommendation solution which synthesizes sentiment aligned neural language explanations to defend its recommendations. The alignment is obtained at the word-level by two customized soft gates, and at the sequence-level by a content-based sentiment regularizer, at both training and inference time.
Offline experiments and user studies demonstrate our model's advantages in both personalized recommendation and explanation tasks.

This work initiates the exploration of the critical role of sentiment in explainable recommendation. It leaves several valuable paths forward. Our sentiment regularizer design enables semi-supervised explainable recommendation via data augmentation.
Considering the extreme sparsity of recommendation data, it can exploit the dominant amount of unobserved data for improved performance. Besides, recommendation eventually is a list-wise ranking problem; thus, it is vital to offer explanations that can reveal the relative order among the recommended items, i.e., a list-wise explanation.

\begin{acks}
We thank the anonymous reviewers for their insightful comments and suggestions. This work is partially supported by the National Science Foundation under grant SCH-1838615, IIS-1553568, and IIS-2007492, and by Alibaba Group through Alibaba Innovative Research Program. 
\end{acks}

\bibliographystyle{ACM-Reference-Format}
\bibliography{main}


\end{document}